\documentclass[aps,prb,twocolumn,superscriptaddress]{revtex4-1}
\usepackage{eurosym}
\usepackage{amsfonts}
\usepackage{amssymb}
\usepackage{amsmath}
\usepackage{graphicx}
\usepackage{color}
\usepackage{chapterbib}
\usepackage{threeparttable}
\usepackage{CJK}
\usepackage{sidecap}
\usepackage[hypertex,dvipdfm,colorlinks=true,urlcolor=blue,linkcolor=blue,citecolor=blue]{hyperref}
\newcommand*{\citen}{}
\DeclareRobustCommand*{\citen}[1]{%
  \begingroup
    \romannumeral-`\x 
    \setcitestyle{numbers}%
    \cite{#1}%
  \endgroup   }
\begin{document}
\title{Inversion symmetry breaking induced triply degenerate points in orderly arranged PtSeTe family materials}

\author{R. C. Xiao}
\email[The authors contributed equally to this work.]{}
\affiliation{Key Laboratory of Materials Physics, Institute of Solid
State Physics, Chinese Academy of Sciences, Hefei 230031, China}
\affiliation{University of Science and Technology of China, Hefei 230026, China}

\author{C. H. Cheung}
\email[The authors contributed equally to this work.]{}
\affiliation{Graduate Institute of Applied Physics, National Taiwan University, Taipei 10617, Taiwan}

\author{P. L. Gong}
\affiliation{Department of Physics, Southern University of Science and Technology, Shenzhen 518055, China}

\author{W. J. Lu}
\email{wjlu@issp.ac.cn}
\affiliation{Key Laboratory of Materials
Physics, Institute of Solid State Physics, Chinese Academy of
Sciences, Hefei 230031, China}

\author{J. G. Si}
\affiliation{Key Laboratory of Materials Physics, Institute of Solid
State Physics, Chinese Academy of Sciences, Hefei 230031, China}
\affiliation{University of Science and Technology of China, Hefei 230026, China}

\author{Y. P. Sun}
\email{ypsun@issp.ac.cn}
\affiliation{Key Laboratory of Materials Physics, Institute of Solid
State Physics, Chinese Academy of Sciences, Hefei 230031, China}
\affiliation{High Magnetic Field Laboratory, Chinese Academy of
Sciences, Hefei 230031, China}
\affiliation{Collaborative Innovation Center of Microstructures,
Nanjing University, Nanjing 210093, China }

\begin{abstract}
$k$ paths exactly with $C_{3v}$ symmetry allow to find triply degenerate points (TDPs) in band structures. The paths that host the type-II Dirac points in PtSe$_2$ family materials also have the $C_{3v}$ spatial symmetry. However, due to Kramers degeneracy (the systems have both inversion symmetry and time reversal symmetry), the crossing points in them are Dirac ones. In this work, based on symmetry analysis, first-principles calculations, and $k\cdot p$ method, we predict that PtSe$_2$ family materials should undergo topological transitions if the inversion symmetry is broken, \emph{i.e.} the Dirac fermions in PtSe$_2$ family materials split into TDPs in PtSeTe family materials (PtSSe, PtSeTe, and PdSeTe) with orderly arranged S/Se (Se/Te). It is different from the case in high-energy physics that breaking inversion symmetry $I$ leads to the splitting of Dirac fermion into Weyl fermions. We also address a possible method to achieve the orderly arranged in PtSeTe family materials in experiments. Our study provides a real example that Dirac points transform into TDPs, and is helpful to investigate the topological transition between Dirac fermions and TDP fermions.
\end{abstract}
\maketitle
\section{Introduction}
The discoveries of Dirac\cite{RN1051,RN1121,RN1123,RN1130,RN1125} and Weyl\cite{RN257,RN1187,RN1659,RN1131,RN1133} semimetals, whose low-energy excitations are analogous to elementary particles in high-energy physics, have made Dirac and Weyl fermions being intensively studied in recent years. The Dirac/Weyl semimetals, in which Dirac/Weyl points are isolated without contacting bulk electron and hole pockets, may correspond to their counterparts in the high-energy physics, are called type-I Dirac/Weyl ones. Besides, unlike high-energy physics, the restriction of Lorentz invariance is not necessary in condensed matter physics. Therefore, many type-II Weyl (e.g. WTe$_2$,\cite{RN762,RN1447} MoTe$_2$,\cite{RN206,RN991,RN1710} Ta$_3$S$_2$,\cite{RN1182} and TaIrTe$_4$\cite{RN1180}) and  type-II Dirac (e.g. PtSe$_2$ family materials,\cite{RN980,RN1580,RN1418,RN1417,RN1581,RN1681} and  VAl$_3$ family materials\cite{RN1448}) materials, in which the cones are seriously titled and crossing points connect bulk electron and hole pockets, have also been discovered. Exotic properties, such as direction dependent chiral anomaly,\cite{RN762,RN1118,RN1183} anti-chiral effect of the chiral Landau level,\cite{RN1184} and novel quantum oscillations\cite{RN1185} also make the type-II Dirac/Weyl semimetals significant compared with the type-I Dirac/Weyl ones.

Whether the band crossings in Driac/Weyl semimetals are type-I or type-II, they are all fourfold/twofold degenerate points. Moreover, condense matters also allow for the existence of other types of unconventional quasi-particle excitations such as three-, six-, or eightfold degenerate points.\cite{RN953,RN1456,RN1457,RN1458,RN1455} The triply degenerate points (TDPs) residing at high symmetry paths in reciprocal spaces result from non-degenerate bands cross double-degenerate ones, so it is necessary for the little groups of $k$ paths to connect one and two dimensional irreducible representations. Hence in most so-far discovered TDP materials (tungsten carbide (WC)-type structure materials,\cite{RN1456,RN1458,RN1455,RN1687} NaCu$_3$Te$_2$,\cite{RN1652,RN1692} simple half-Heusler,\cite{RN1688} and metal diborides\cite{RN1650}), the $k$ paths that host TDPs mostly have ${{C}_{3v}}$ symmetry.

TDPs in the condensed matters can be viewed as intermediate topological phases between Dirac points and Weyl points. PtSe$_2$ family materials (PtSe$_2$, PtTe$_2$, and PdTe$_2$) with $D_{3d}$ point group symmetry (Fig. \ref{crystal}(a)) were recently predicted by theories\cite{RN980} and confirmed in experiments to be type-II topological Dirac materials.\cite{RN1580,RN1418,RN1417,RN1581,RN1681} $\Gamma -A$ ($\Delta $) paths that host the type-II Dirac points in PtSe$_2$ family materials also have the ${{C}_{3v}}$ spatial symmetry. However, the materials have both inversion symmetry $I$ and time reversal symmetry $T$. Thus the two non-degenerate bands are degenerate due to Kramers degeneracy. If the inversion symmetry $I$ is broken, the Kramers degeneracy along $\Gamma -A$ path is generally broken, and TDPs can exist under this symmetry condition. It is natural to consider the chalcogen substitution in PtSe$_2$ family materials to break the inversion symmetry $I$. In this work, we focus on the PtSSe, PtSeTe, and PdSeTe, which have been previously synthesized in experiments.\cite{RN1654,RN1653} In order to simplify calculation and catch main physics, we further consider the orderly arranged ones as shown in Fig. \ref{crystal}(b). Therefore, the little groups of $\Gamma -A$ ($\Delta $) path merely has the ${{C}_{3v}}$ spatial symmetry without $T\cdot I$ symmetry.

In the following main text, taking PtSeTe as a representation, we propone that the Dirac fermions in PtSe$_2$ family materials split into TDPs in orderly arranged PtSeTe family materials (PtSSe, PtSeTe, and PdSeTe) by breaking inversion symmetry $I$, and we analyses the mechanism of the topological transitions in two family materials. Furthermore, the phonon stabilities of PtSeTe family materials are studied, and a possible synthesis method is also addressed.

\section{Calculation details}
The first-principles calculations based on density functional theory (DFT) were performed using QUANTUM-ESPRESSO package.\cite{RN82} Ultrasoft pseudo-potentials and general gradient approximation (GGA) according to the PBE functional were used. The energy cutoff of the plane wave (charge density) basis was set to 50 Ry (500 Ry). The Brillouin zone was sampled with a 12$\times $12$\times $8 $k$-points mesh. The lattice constants and ion positions were optimized using the Broyden-Fletcher-Goldfarb-Shanno (BFGS) quasi-Newton algorithm. All of band structure calculations were cross-checked by VASP codes,\cite{RN1434,RN1433} and the results are consistent with each other. Phonon spectra were calculated using density functional perturbation theory (DFPT)\cite{RN131} with a 6$\times $6$\times $4 $q$-points mesh. Low-energy effective Hamiltonian were studied by $k\cdot p$ method with invariant theory.
\section{Results and Discission}
\begin{figure}
\centering
\includegraphics[width=1\columnwidth]{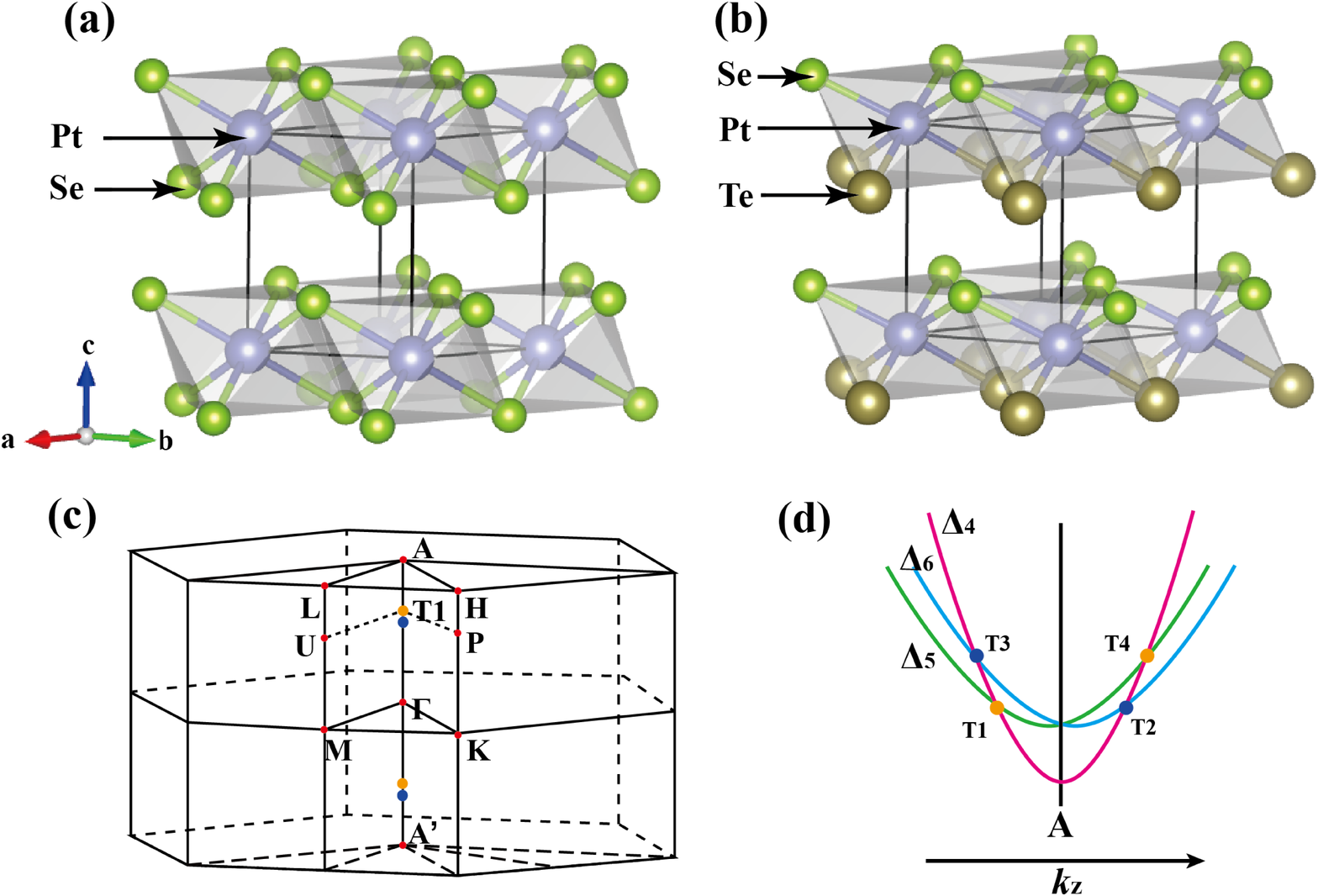}
\caption{Crystal structures of (a) PtSe$_2$ and (b) PtSeTe. (c) Brillouin zone of PtSeTe. $A'$ is equivalent to $A$ point, and lies beneath $\Gamma $ point. (d) Illustration of TDPs near $A$ point. Orange and blue dots in (c) and (d) stand for TDPs originating from ${{\Delta }_{4}}$ band crossing ${{\Delta }_{5}}$ and ${{\Delta }_{6}}$ bands, respectively.}
\label{crystal}
\end{figure}

\begin{figure*}
\centering
\includegraphics[width=0.98\textwidth]{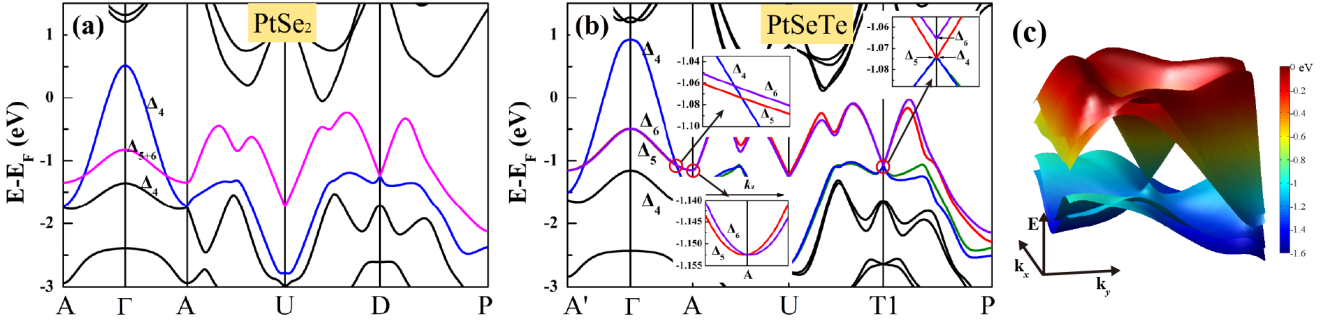}
\caption{(a) Band structure of PtSe$_2$. ${{\Delta }_{5+6}}$ means degenerate ${{\Delta }_{5}}$ and ${{\Delta }_{6}}$ bands, and D point denotes position where Dirac cone locates along  $\Gamma -A$ path. (b) Band structure of PtSeTe. Insets are zoomed in pictures near $A$ and $T1$ points. (c) Three dimensional band structure of PtSeTe on ${{k}_{x}}-{{k}_{y}}$ plane of ${{k}_{z}}=0.370$ ($-0.2\le {{k}_{x}},{{k}_{y}}\le 0.2$, all in crystal coordinate).}
\label{bandstr}
\end{figure*}

PtSe$_2$, PtTe$_2$, and PdTe$_2$ belong to transition metal dichalcogenides with ${{D}_{3d}}$ point group and time reversal symmetry $T$. Each has a layered structure with one transition metal atom and two X (X=Se, Te) atoms located at (0,0,0), $\left( {1}/{3}\;,{2}/{3}\;,{{z}_{\text{X}}} \right)$, and $\left( {2}/{3}\;,{1}/{3}\;,-{{z}_{\text{X}}} \right)$ sites, respectively (Fig. \ref{crystal}(a)). The band structure of PtSe$_2$ with spin orbit coupling (SOC) effects is shown in Fig. \ref{bandstr}(a). Because $\Gamma -A$ path processes the ${{C}_{3v}}$ and $T\cdot I$ symmetries, ${{\Delta }_{5}}$ (basis: $\left( \left|3/2,-3/2\right\rangle -i\left|3/2,3/2\right\rangle  \right)$) and ${{\Delta }_{6}}$ (basis: $\left( i\left|3/2,-3/2\right\rangle -\left|3/2,3/2\right\rangle  \right)$)\cite{RN1695} bands are degenerate (Kramers degeneracy, denoted as ${{\Delta }_{5+6}}$), and they cross double-degenerate ${{\Delta }_{4}}$ (basis: $\left( \left| 1/2,1/2 \right\rangle ,\left| 1/2,-1/2 \right\rangle  \right)$) band, forming the Dirac points. Therefore, the Dirac points in PtSe$_2$ are protected by ${{C}_{3v}}\otimes {\bar{1}}'$ symmetry (${\bar{1}}'\text{=}\left\{ \mathbb{I},T\cdot I \right\}$, $\mathbb{I}$: identity). Furthermore, the Dirac cones are tilted along $\Gamma -A$ path, which connect the electronic and hole pockets, and called type-II Dirac cones. Due to similar crystal structure, the band structures of PtSe$_2$, PtTe$_2$, and PdTe$_2$ resemble each other.\cite{RN1681,RN1460,RN980} Though the Dirac points are far away from Fermi energy, theory\cite{RN1580} and experiment\cite{RN1647} have shown that it is possible to tune the Dirac points to the Fermi energy by doping Ir in PtTe$_2$ (\emph{i.e.} Pt$_{1-x}$Ir$_{x}$Te$_2$). Tight-binding mode predicted that the type-II Dirac cones can be transformed into type-I ones by controlling inter-layer hopping.\cite{RN1681} DFT calculations also revealed that external pressure can manipulate the type-II and type-I Dirac points in PdTe$_2$, PtTe$_2$, and PdTe$_2$.\cite{RN1460}

\begin{table}[!htbp]
\centering
\caption{ Experimental and optimized lattice constants of PtSSe, PtSeTe, and PdSeTe (in $\text{\AA}$)}
\setlength{\tabcolsep}{5mm}{
\begin{tabular}{c|ccc}
   \hline
   \hline
           &     PtSSe  & PtSeTe &	PdSeTe    \\
    \hline
${{a}_{Expt.}}$  & 3.59 &	3.89 &	3.90                           \\
${{c}_{Expt.}}$  & 5.06 &	5.11 &	4.98                          \\
    \hline	
${{a}_{Opt.}}$   & 3.66 &	3.91 &	3.96                          \\
${{c}_{Opt.}}$   & 5.04 &	5.07 &	4.98                         \\
\hline
\hline
\end{tabular}}
\begin{tablenotes}
\item[1] Note: Experimental data are from Refs. ~\citen{RN1654} and ~\citen{RN1653}.
\end{tablenotes}
\label{Tab1}
\end{table}

When one Se layer in PtSe$_2$ is replaced by one Te layer, and Se and Te layers are and orderly arranged as shown in Fig. \ref{crystal} (b), then the inversion symmetry $I$ is absent in PtSeTe, so its spatial symmetry is reduced from ${{D}_{3d}}$ to ${{C}_{3v}}$ (${{D}_{3d}}={{C}_{3v}}\otimes {{C}_{i}}$, ${{C}_{i}}=\left\{ \mathbb{I},I \right\}$). The optimized lattice constants of PtSeTe are in agreement with the experiments as shown in Table I. $\Gamma-A$ path still has the ${{C}_{3v}}$ spatial symmetry, but no $T\cdot I$ symmetry. In this case, the ${{\Delta }_{5}}$ and ${{\Delta }_{6}}$ bands are not degenerate at all, then the split ${{\Delta }_{5}}$ and ${{\Delta }_{6}}$ bands cross ${{\Delta }_{4}}$ band, forming two TDPs, as shown in Fig. \ref{bandstr}(b). Near the TDPs, the splitting energy of the ${{\Delta }_{5}}$ and ${{\Delta }_{6}}$ bands ($\Delta E$) is about 9 meV. As expected, there is another pair of TDPs along $\Gamma -A'$ line (Fig. \ref{bandstr}(b)). The position of these two pairs of TDPs in the reciprocal space are $\left( 0,0,\pm 0.368 \right)$ and $\left( 0,0,\pm 0.370 \right)$ (in crystal coordinate), respectively. While at the $\Gamma $ and $A$ points, the ${{\Gamma }_{5}}$ and ${{\Gamma }_{6}}$ states are degenerate again, due to possessing the $T$ symmetry at these points (Fig. \ref{bandstr}(b) inset). The splitting energy of the ${{\Delta }_{5}}$ and ${{\Delta }_{6}}$ bands is small along $\Gamma -A$ line, but the splitting of the two band surface can be a large value  on ${{k}_{x}}-{{k}_{y}}$ plane (Fig. \ref{bandstr}(c)), so the separation on this plane of these two bands is easier to be detected by experiments. ${{\Delta }_{4}}$ band surfaces are also split on this plane (Fig. \ref{bandstr}(c)), also due to Kramers degeneracy breaking.

In high-energy physics, breaking time reversal $T$ or inversion symmetry $I$ can lead to the splitting of Dirac fermion into Weyl fermions. However, in condensed matter physics, fermions are constrained by the crystal symmetries rather than by the Lorentz invariance. The symmetry group of any $k$ path of any symmorphic system can be described by magnetic point group. ${{C}_{3v}}\otimes {\bar{1}}'$ can be classified as the black and white magnetic point group of ${{\overline{3}}^{\prime }}m$.\cite{RN1704} Therefore, though PtSe$_2$ family materials are non-magnetic, their $\Gamma -A$ paths can be labeled by black and white magnetic point group ${{\overline{3}}^{\prime }}m$. Anti-unitary operator $T\cdot I$ makes irreducible representations ${{\Delta }_{5}}$ and ${{\Delta }_{6}}$ of ${{C}_{3v}}$ degenerate. The symmetries of $\Gamma -A$ paths in orderly arranged PtSeTe family materials belong to original point group $3m$ (${{C}_{3v}}$), without anti-unitary $T\cdot I$ symmetry. So the magnetic point groups of the $\Gamma -A$ path of two kinds of material are different.

\begin{figure}
\centering
\includegraphics[width=0.9\columnwidth]{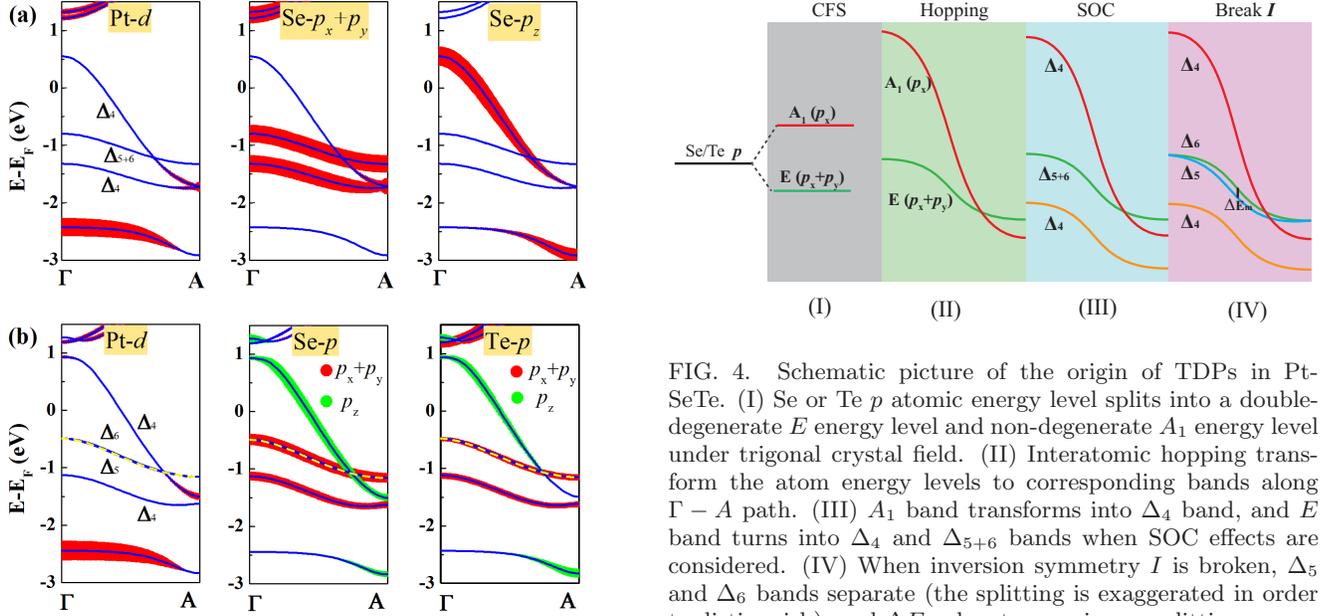}
\caption{Orbital resolved bands (fat-bands) of (a) PtSe$_2$ and (b) PtSeTe along $\Gamma -A$ path.}
\label{fatband}
\end{figure}

The orbital resolved bands of PtTe$_2$ along the $\Gamma -A$ path are shown Fig. \ref{fatband}(a). The upper ${{\Delta }_{4}}$ band is mainly composed by Se-${{p}_{z}}$ orbits, while the ${{\Delta }_{5+6}}$ and the lower ${{\Delta }_{4}}$ bands are mainly composed by Se-${{p}_{x}}+{{p}_{y}}$ orbits. So is PtSeTe, as shown in Fig. \ref{fatband}(b). Therefore, the bands forming the Dirac points or TDPs are mainly composed by Se/Te-$p$ orbits. Additionally, both for PtSe$_2$ and PtSeTe, Pt-$d$ and Se/Te-$p$ orbits hybrid in the ${{\Delta }_{5}}$ and ${{\Delta }_{6}}$ bands near the $A$ points.

\begin{figure}
\centering
\includegraphics[width=0.98\columnwidth]{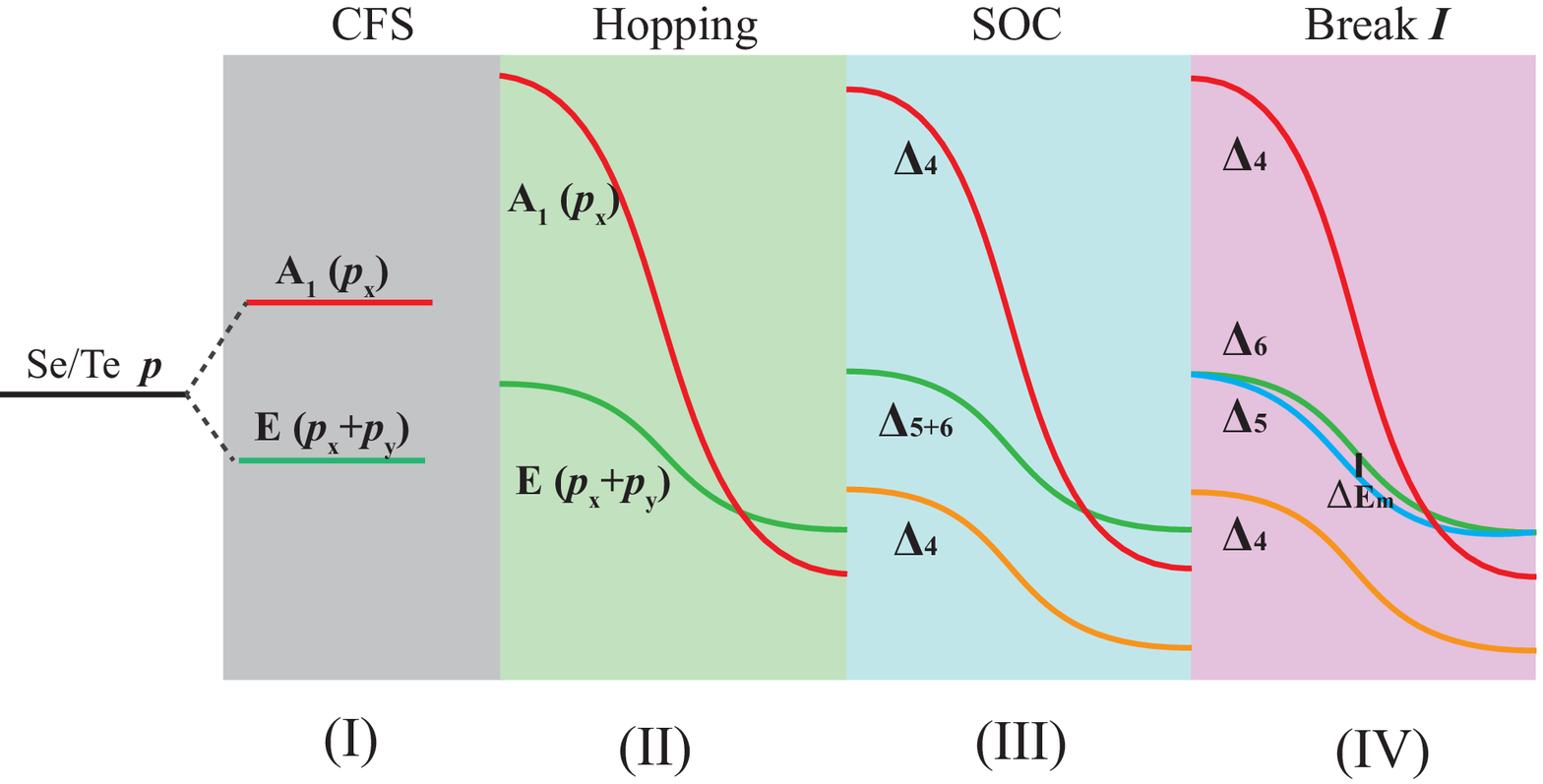}
\caption{Schematic picture of the origin of TDPs in PtSeTe. (I) Se or Te $p$ atomic energy level splits into a double-degenerate $E$ energy level and non-degenerate ${{A}_{1}}$ energy level under trigonal crystal field. (II) Interatomic hopping transform the atom energy levels to corresponding bands along $\Gamma -A$ path. (III) ${{A}_{1}}$ band transforms into ${{\Delta }_{4}}$ band, and $E$ band turns into ${{\Delta }_{4}}$ and ${{\Delta }_{5+6}}$ bands when SOC effects are considered. (IV) When inversion symmetry $I$ is broken, ${{\Delta }_{5}}$ and ${{\Delta }_{6}}$ bands separate (the splitting is exaggerated in order to distinguish), and $\Delta {{E}_{m}}$ denotes maximum splitting energy of two bands.}
\label{Evaluation}
\end{figure}

Starting from the atomic orbits of Se/Te, the following four steps are required to understand TDPs in PtSeTe, as illustrated in Fig. \ref{Evaluation}. Se/Te atoms locate in the trigonal crystal field, and every $p$ atomic energy level transforms into a double-degenerate $E$ energy level (from ${{p}_{x}}$ and ${{p}_{y}}$ orbits) and non-degenerate ${{A}_{1}}$ energy level (from ${{p}_{z}}$ orbits) under crystal field splitting (CFS) as shown in Fig. \ref{Evaluation}(I). The interatomic hopping transforms the atom energy levels into ${{A}_{1}}$ and $E$ bands along the $\Gamma -A$ path. The out-of-plane ${{p}_{z}}$ orbits usually have much larger the hopping than in-plane ${{p}_{x}}$ and ${{p}_{y}}$ orbits along the $\Gamma -A$ path,\cite{RN1681} so the ${{A}_{1}}$ band is more dispersive than the $E$ one, and they cross each other as shown in Fig. \ref{Evaluation}(II). When SOC is considered, the ${{A}_{1}}$ band transforms into the ${{\Delta }_{4}}$ one, and the $E$ band transforms into ${{\Delta }_{4}}$ and ${{\Delta }_{5+6}}$ bands (see the Supplemental Material\cite{RN1695}). Furthermore, the ${{\Delta }_{4}}$ band originating from ${{A}_{1}}$ band crosses the ${{\Delta }_{5+6}}$ band forming a type-II Dirac point in PtSe$_2$ or PtTe$_2$ (Fig. \ref{Evaluation}(III)). When the inversion symmetry is broken, the ${{\Delta }_{5+6}}$ band splits into ${{\Delta }_{5}}$ and ${{\Delta }_{6}}$ ones, and one Dirac point transforms into two TDPs as shown in Fig. \ref{Evaluation}(IV). So the TDPs in PtSeTe can be explained by the evolution of Se/Te-$p$ orbital energy levels as a result of CFS, interatomic hopping, SOC effects, and inversion symmetry breaking.

To further investigate the nature of TDPs in PtSeTe, we constructed a low-energy effective Hamiltonian using $k\cdot p$ method. The little group of $A$ point is ${{C}_{3v}}\otimes {1}'$ (${1}'=\{\mathbb{I},T\}$). The $4\times 4$ effective Hamiltonian (bases are in the order of $\left\{ \Delta_{4},\Delta_{5},\Delta_{6} \right\}$) that considers first-order $k$ terms for off-diagonal matrix elements and second-order $k$ terms for diagonal matrix elements is (details are in the Supplemental Material\cite{RN1695})
\begin{equation}
\begin{aligned}
& {{H}_{eff}}(\mathbf{q})={{\varepsilon }_{0}}(\mathbf{q}) + \\
&  \left[ \begin{matrix}
   M(\mathbf{q}) & iC{{q}_{+}} & A{{q}_{-}} & -{{A}^{*}}{{q}_{-}}  \\
   -iC{{q}_{-}} & M(\mathbf{q}) & -A{{q}_{+}} & -{{A}^{*}}{{q}_{+}}  \\
   {{A}^{*}}{{q}_{+}} & -{{A}^{*}}{{q}_{-}} & -M(\mathbf{q})+B{{q}_{z}} & 0  \\
   -A{{q}_{+}} & -A{{q}_{-}} & 0 & -M(\mathbf{q})-B{{q}_{z}}  \\
\end{matrix} \right]
\end{aligned}
\label{Heff}
\end{equation}
where $\varepsilon_{0}(\mathbf{q})={{C}_{0}}+{{C}_{1}}q_{z}^{2}+{{C}_{2}}(q_{x}^{2}+q_{y}^{2})$, $M(\mathbf{q})={{M}_{0}}-{{M}_{1}}q_{z}^{2}-{{M}_{2}}(q_{x}^{2}+q_{y}^{2})$, $\mathbf{q}=\mathbf{k}-\mathbf{A}$, ${{q}_{+}}={{q}_{x}}+i{{q}_{y}}$, and ${{q}_{-}}={{q}_{x}}-i{{q}_{y}}$. $A$ is a complex number, while $B$ and $C$ are real numbers. $B{{q}_{z}}$ exists in ${{H}_{33}}$ and ${{H}_{44}}$, and this term makes ${{\Delta }_{5}}$ and ${{\Delta }_{6}}$ bands spilt. We can get the eigenvalues along the $\Gamma -A$ path:
\begin{eqnarray}
&&  {{E}_{\Delta 4}}={{C}_{0}}+{{C}_{1}}q_{z}^{2}+{{M}_{0}}-{{M}_{1}}q_{z}^{2}, \label{delta4}   \\
&&  {{E}_{\Delta 5}}={{C}_{0}}+{{C}_{1}}q_{z}^{2}-{{M}_{0}}+{{M}_{1}}q_{z}^{2}+B{{q}_{z}}, \label{delta5}   \\
&&  {{E}_{\Delta 6}}={{C}_{0}}+{{C}_{1}}q_{z}^{2}-{{M}_{0}}+{{M}_{1}}q_{z}^{2}-B{{q}_{z}}. \label{delta6}
\end{eqnarray}
According to Eqs. \ref{delta5} and \ref{delta6}, ${{\Delta }_{5}}$ and ${{\Delta }_{6}}$ states are degenerate at $A$ point (${{q}_{z}}=0$, \emph{i.e.} ${{A}_{5}}$ and ${{A}_{6}}$), coinciding with the DFT results. The ${{\Delta }_{4}}$ and ${{\Delta }_{5}}$ bands cross at ${{q}_{z}}={\left( -B\pm \sqrt{{{B}^{2}}+16{{M}_{0}}{{M}_{1}}} \right)}/{4{{M}_{1}}}$, while the ${{\Delta }_{4}}$ and ${{\Delta }_{6}}$ bands cross at ${{q}_{z}}={\left( B\pm \sqrt{{{B}^{2}}+16{{M}_{0}}{{M}_{1}}} \right)}/{4{{M}_{1}}}$. Using the $k\cdot p$ effective Hamiltonian around the TDPs, we can prove that each TDP can be regard as two Weyl points with opposite chirality (see the Supplemental Material\cite{RN1695}). If the mirror symmetry of PtSeTe is broken (such as applying magnetic field along $z$ direction), the symmetry of $\Gamma -A$ path transforms from ${{C}_{3v}}$ into ${{C}_{3}}$, and ${{H}_{11}}$ and ${{H}_{22}}$ will also process linear ${{q}_{z}}$ terms. Hence, the ${{\Delta }_{4}}$ band splits, and each TDP transforms into two Weyl points when the mirror symmetry is broken.

\begin{figure}
\centering
\includegraphics[width=0.8\columnwidth]{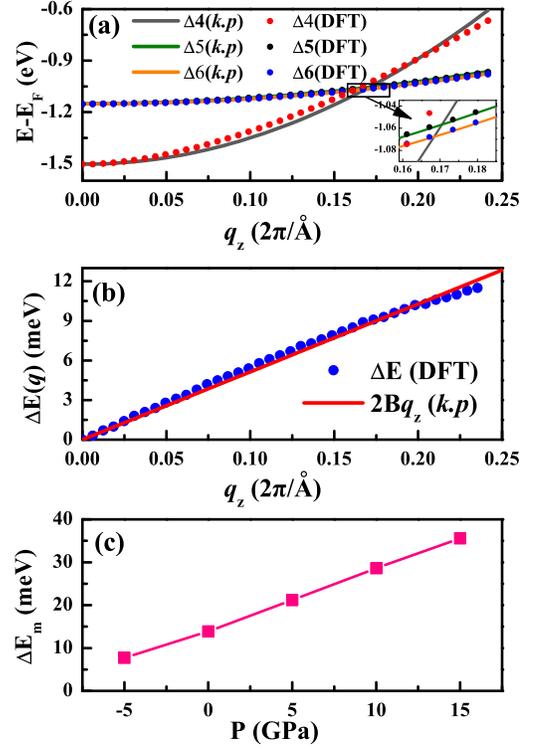}
\caption{ (a) Bands along $\Gamma -A$ path for $k\cdot p$ model compared with DFT results. (b) $\Delta E(q)$ near the $A$ point. (c) Maximum splitting energy $\Delta {{E}_{m}}$ with pressure (obtained by DFT calculations).}
\label{deltaE}
\end{figure}

\begin{figure}
\centering
\includegraphics[width=0.87\columnwidth]{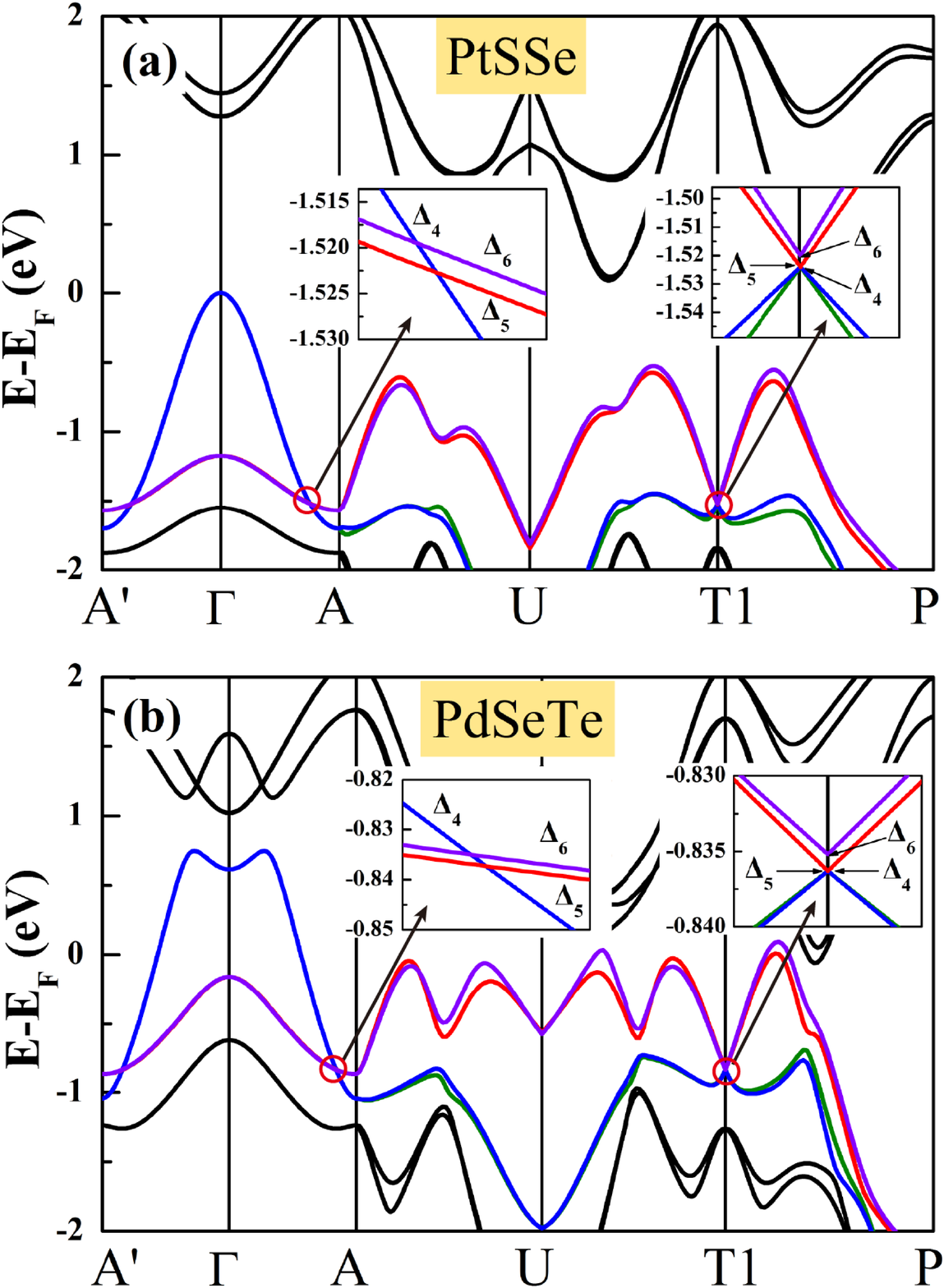}
\caption{Band structures of (a) PtSSe and (b) PdSeTe.}
\label{PtSSe-PdSeTe}
\end{figure}

The $k\cdot p$ model bands along the $\Gamma -A$ path are shown in Fig. \ref{deltaE}(a). According to Eqs. \ref{delta5} and \ref{delta6}, the splitting energy of ${{\Delta }_{5}}$ and ${{\Delta }_{6}}$ bands $\Delta E(q)=2B{{q}_{z}}$ (see Fig. \ref{deltaE}(b)), so $B$ determines the splitting between the ${{\Delta }_{5}}$ and ${{\Delta }_{6}}$ bands. $\Delta E$ originates from the second-order interaction between the ${{J}_{z}}=\pm 3/2$ (${{\Delta }_{5}}$ and ${{\Delta }_{6}}$) states and the metal cation (Pt) $d$ core levels.\cite{RN1652,RN1694} According to Ref. ~\citen{RN1652}, $\Delta E(q)$ can be changed by varying $p-d$ hybridization of metal and anion atoms. To reveal the effect of $p-d$ hybridization on the splitting, we perform a hypothetical experiment that applying positive and negative pressure on PtSeTe to change the lattice constants by using DFT calculations, and we find that positive (negative) pressure does increase (decrease) $\Delta {{E}_{m}}$ (see Fig. \ref{deltaE}(c)).

\begin{figure}[t]
\centering
\includegraphics[width=0.87\columnwidth]{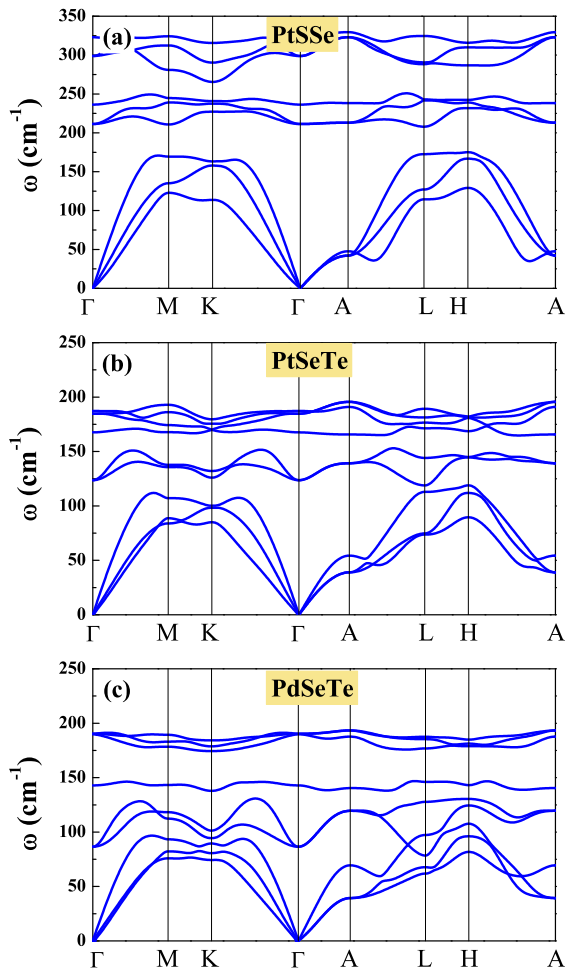}
\caption{ Phonon spectra of (a) PtSSe, (b) PtSeTe, and (c) PdSeTe.}
\label{phonon}
\end{figure}

Having examined PtSeTe, now we move on to PtSSe and PdSeTe. Optimized lattice constants of PtSSe, and PdSeTe are shown in Table I, experimental ones are also given, and they coincide with each other. We find that the Dirac points in PtSe$_2$ and PdTe$_2$ also transform into the TDPs in orderly arranged PtSSe and PdSeTe (see Fig. \ref{PtSSe-PdSeTe}). The calculated phonon spectra of these three orderly arranged compounds are all stable as shown in Fig. \ref{phonon}. Previous study have shown that S and Se can be orderly arranged in 1$T$-TaS$_{2-x}$Se$_{x}$.\cite{RN1550} Mono to several layers high-quality PtSe$_2$ samples have been grown by molecular beam epitaxy (MBE) method.\cite{RN1701,RN1702} Therefore, MBE could be a promising method to synthesize the orderly arranged PtSeTe, PtSSe, and PdSeTe by one-by-one atomic layer growing sequence.

\section{Conclusion}
Due to ${{C}_{3v}}\otimes {\bar{1}}'$ symmetry, ${{\Delta }_{5}}$ and ${{\Delta }_{6}}$ bands degenerate in PtSe$_2$ family materials. By breaking $I$ symmetry, Kramers degenerate ${{\Delta }_{5}}$ and ${{\Delta }_{6}}$ bands split, therefore each Dirac cone in PtSe$_2$ family materials transform into two TDPs in orderly arranged PtSeTe, PtSSe, and PdSeTe where the $\Gamma -A$ paths only preserve ${{C}_{3v}}$ symmetry. Unlike the high-energy physics, breaking the inversion symmetry $I$ of PtSe$_2$ family materials leads to the splitting of Dirac fermion into TDPs rather than Weyl fermions. The splitting energy $\Delta E$ of ${{\Delta }_{5}}$ and ${{\Delta }_{6}}$ bands can be manipulated by $p-d$ hybridization of metal and anion atoms. These three materials are stable in phonon spectra and very possible to be grown in experiments, such as using MBE method. If orderly arranged PtSeTe family materials are synthesized, they will provide real examples, in which Dirac points transform to TDPs points, and can help us to investigate the topological transition between Dirac fermions and TDP fermions. Further experimental verification and theoretical studies need to be carry out.

\begin{acknowledgements}
This work is supported by the National Key R\&D Program of China under Contract No. 2016YFA0300404, the National Nature Science Foundation of China under Contracts Nos. 11674326, 11774351, 11704001, and U1232139. The calculations were partially performed at the Center for Computational Science, CASHIPS.

R.C.X carried out all the calculations and wrote the paper with assistance from P.L.G. and J.G.S.. C.H.C. performed the symmetry analysis and checked $k\cdot p$ calculation and paper writing. W.J.L. planned and integrated the research, and revised the manuscript. Y.P.S. supervised the project.
\end{acknowledgements}

\end {document}